\documentclass[prb,aps,twocolumn,superscriptaddress,showpacs,nofootinbib]{revtex4-2}

\usepackage{graphicx}
\DeclareUnicodeCharacter{2009}{\,} 
\DeclareUnicodeCharacter{2212}{--}
\DeclareUnicodeCharacter{2A7D}{$\leq$}
\DeclareUnicodeCharacter{03B2}{$\beta$}
\usepackage{dcolumn}
\usepackage{bm}
\usepackage[mathlines]{lineno}
\usepackage{booktabs} 

\renewcommand{\selectlanguage}[1]{} 

\usepackage[utf8]{inputenc}
\usepackage[T1]{fontenc}
\usepackage{mathptmx}
\usepackage{etoolbox}
\usepackage{float}
\usepackage{xcolor}

\begin{document}

\title{Magnetic dilution in the triangular lattice antiferromagnet NaYb$_{1-x}$Lu$_{x}$O$_2$}

\author{Steven J. Gomez Alvarado}
\affiliation{Materials Department, University of California, Santa Barbara, CA, USA}
\author{Brenden R. Ortiz}
\affiliation{Materials Science and Technology Division, Oak Ridge National Laboratory, Oak Ridge, TN, USA}
\author{Soren Bear}
\affiliation{Materials Department, University of California, Santa Barbara, CA, USA}
\author{Benito A. Gonzalez}
\affiliation{Materials Department, University of California, Santa Barbara, CA, USA}
\author{Andrea N. Capa Salinas}
\affiliation{Materials Department, University of California, Santa Barbara, CA, USA}
\author{Adam Berlie}
\affiliation{ISIS Neutron and Muon Source, STFC, Rutherford Appleton Laboratory, Chilton, Didcot, Oxon OX11 0QX, UK}
\author{Michael J. Graf}
\affiliation{Department of Physics, Boston College, Chestnut Hill, MA, USA}
\author{Stephen D. Wilson}
\altaffiliation{stephendwilson@ucsb.edu}
\affiliation{Materials Department, University of California, Santa Barbara, CA, USA}

\date{\today}

\begin{abstract}

{The delafossite-like compound NaYbO$_2$ hosts a triangular lattice of Yb$^{3+}$ moments and is a promising candidate for the realization of a quantum spin liquid ground state -- an exotic, quantum-disordered magnetic phase featuring long-range entanglement of spins. Tuning this system away from this quantum-disordered regime toward classical order or spin freezing is a powerful approach to shed light on the nature of the parent ground state. Here we leverage the substitution of nonmagnetic Lu$^{3+}$ onto the Yb$^{3+}$ sites to study the effects of magnetic disorder in NaYbO$_2$ using low-temperature ac susceptibility, heat capacity, and muon spin relaxation ($\mu$SR) measurements. Our $\mu$SR measurements reveal resilient, correlated magnetic fluctuations that persist to at least 15\% dilution, precluding conventional spin freezing and magnetic inhomogeneity. Heat capacity and magnetic susceptibility resolve a rapid suppression of the field-induced ``up-up-down'' magnetic order upon dilution and a crossover in the power-law behavior of the low-temperature magnetic excitations associated with the zero-field quantum disordered ground state. Taken together, these results support the notion of a robust network of entangled moments in NaYbO$_2$, and provides experimental validation of several models of a Heisenberg triangular lattice antiferromagnet in the presence of disorder.}

\end{abstract}

\maketitle

\section{Introduction}

Quantum spin liquid (QSL) states are a theorized state of matter where highly entangled spin states result in interesting quantum phenomena, notably the formation of many-body entangled states and fractionalized spin excitations \cite{savary_quantum_2016}.
The compound NaYbO$_2$ features a perfect triangular lattice of $S_\mathrm{eff}=1/2$ moments, which show no signs of spin freezing or glassiness down to 50~mK, making this system a robust platform for exploring spin liquid physics \cite{bordelon_field-tunable_2019, PhysRevB.99.180401, PhysRevB.100.144432}. NaYbO$_2$ is proposed to host an intrinsic quantum disordered ground state proximate to magnetic order.  This is demonstrated by a zero field ground state where the Yb$^{3+}$ moments are dynamic and where modest applied fields are shown to drive collinear ordering into an up-up-down state. This field-driven instability is naively more sensitive to disorder than an otherwise hidden non-collinear zero-field ground state \cite{bordelon_field-tunable_2019}. The cumulative picture is one of an underlying \textit{XXZ} magnetic Hamiltonian with slight \textit{XY} exchange anisotropy and an easy-plane g-tensor anisotropy that can be field-tuned from quantum disordered into classically ordered states. 

Tuning this model QSL candidate system across phase boundaries from quantum disorder into classical order/freezing stands to shed light on the microscopic mechanism responsible for its magnetic behavior. One way to tune the compound across this phase boundary is via magnetic dilution: replacing magnetic Yb$^{3+}$ ions with nonmagnetic ions, which perturbs quantum fluctuations by introducing local magnetic disorder. For instance, magnetic dilution leveraged in other compounds provided insights into the nature of local spin correlations \cite{PhysRevLett.103.047201, doi:10.1126/science.1067110, PhysRevMaterials.6.046201}, and is proposed as means of differentiating systems with so-called ``confined'' (\textit{i.e.}, conventional) and ``deconfined'' (\textit{i.e.}, fractionalized) magnetic excitations or quasiparticles \cite{sachdev_non-magnetic_2000}. 

Several theories exist regarding the expected ground state when adding disorder to an entangled network of moments on a triangular lattice -- if the parent ground state is a network of entangled singlet states, disorder is predicted to break local singlet pairings and induce the least collinear antiferromagnetic order by creating local spin textures around each impurity \cite{maryasin_triangular_2013, nagaosa_nonmagnetic_1996}. Alternatively, other studies suggest that long-range entanglement is robust, and impurities do not induce local moments \cite{sachdev_non-magnetic_2000, dommange_static_2003}.

Muon spin relaxation ($\mu$SR) studies show similar spin behaviors for NaYbO$_2$ \cite{PhysRevB.100.144432}, NaYbS$_2$ \cite{PhysRevB.100.241116}, and NaYbSe$_2$ \cite{PhysRevB.106.085115, zhu_fluctuating_2023} as they are cooled toward their ground states.  The spin depolarization is generally characterized by the lack of magnetic order in zero magnetic field (down to mK with $T\ll T_\mathrm{CW}$) and by the onset of an enhanced depolarization rate due to coherent magnetic fluctuations below $\approx10$~K . Refs. \cite{PhysRevB.100.144432} and \cite{zhu_fluctuating_2023} report two channels of depolarization at low temperature, manifest as two distinct depolarization rates. The interpretation of the two responses in the low temperature ground state remains debated. Whereas in Ref. \cite{PhysRevB.100.144432} the two components are attributed to muons stopping at crystallographically distinct sites, the authors of Ref. \cite{zhu_fluctuating_2023} instead propose that NaYbSe$_2$ is intrinsically, magnetically inhomogeneous below 10~K. Nuclear magnetic resonance (NMR) measurements on NaYbSe$_2$, however, show no evidence for magnetic inhomogeneity \cite{PhysRevB.100.224417, PhysRevB.110.214440}. Due to its unique, dual sensitivity to spin dynamics in distinct volume fractions of multiphase mixtures, $\mu$SR measurements of magnetic dilution of triangular lattice Yb$^{3+}$ moments stand to provide significant insights into the impact of disorder on the intrinsic quantum disorder manifest in $A$Yb$X_2$ compounds.

 Due to its stronger O-based ligand field, NaYbO$_2$, in particular, stands out in the $A$Yb$X_2$ family with a well-isolated $S_\mathrm{eff}=1/2$ doublet ground state and large charge gap. It is an exceptionally clean material manifestation of intrinsic quantum disorder from which to probe the impact of magnetic impurities within a putative spin liquid ground state. Substitution of nonmagnetic Lu$^{3+}$ for Yb$^{3+}$ would be expected to significantly alter the formation of any inhomogeneous ground state, and breaking bonds within an extended network of singlets allows for tests of the resilience of the QSL phase. 

Here we report $\mu$SR measurements of the magnetically diluted QSL-like state in NaYb$_{1-x}$Lu$_x$O$_2$ with $x=0.05$ and $x=0.15$, as well as supporting magnetic and thermodynamic measurements for $0 \leq x \leq 0.50$. Consistent with earlier results, we find the quantum disordered ground state to be surprisingly robust and insensitive to magnetic dilution, whereas the field-induced collinear up-up-down state is rapidly suppressed. $\mu$SR data show persistent spin dynamics nearly identical to that found for NaYbO$_2$ \cite{PhysRevB.100.144432} with no evidence for magnetic inhomogeneity, whereas the power law associated with the low-frequency excitations sampled via heat capacity switches from a Dirac QSL-like $T^2$ to a linear-$T$ behavior. Our results are interpreted assuming a dynamic, magnetically uniform ground state characterized by a power-law cutoff autocorrelation function with a low-frequency spectrum that is shifted upon dilution of the triangular lattice network.

\section{Methods}

\subsection{Synthesis}

Polycrystalline NaYb$_{1-x}$Lu$_x$O$_2$ samples were prepared by solid-state reaction \cite{bordelon_field-tunable_2019}. Powders were synthesized at compositions $x = \{0.00,0.05,0.10,0.15,0.25,0.50,1.00\}$. Precursors of Yb$_2$O$_3$ (99.99\% metals basis, Sigma Aldrich) and Lu$_2$O$_3$ (99.99\%, Thermo Fisher) were first reacted at 1,000~$^\circ$C for 12~hours in air. This mixture was then combined with Na$_2$CO$_3$ (99.997\%, Alfa Aesar) in a 1:1.25~molar ratio to account for vapor phase losses of Na during the subsequent reaction. While grinding this mixture, the mortar and pestle were cleaned thoroughly before a second grinding step to avoid residual (Yb,Lu)$_2$O$_3$ after the reaction. Once homogenized, the powder was reacted at 1,000~$^\circ$C for three days, with subsequent regrinding and reheating to 1,000~$^\circ$C for another day. All samples were stored in a dry, inert atmosphere and minimally exposed to atmospheric conditions before measurements. 

The phase-purity of samples was verified using laboratory powder X-ray diffraction measurements, carried out using a Panalytical Empyrean diffractometer using Cu--K$\alpha$ radiation. X-ray diffraction measurements showed no evidence of secondary phases (\textit{e.g.}, (Yb,Lu)$_2$O$_3$ or Na$_2$CO$_3$) within the resolution of the diffractometer. {Composition was verified using energy-dispersive X-ray spectroscopy (EDS), confirming good agreeement between nominal and actual Lu stoichiometry (Fig. S1).}

\begin{figure}[t]
    \centering
    \includegraphics[width=1.0\linewidth]{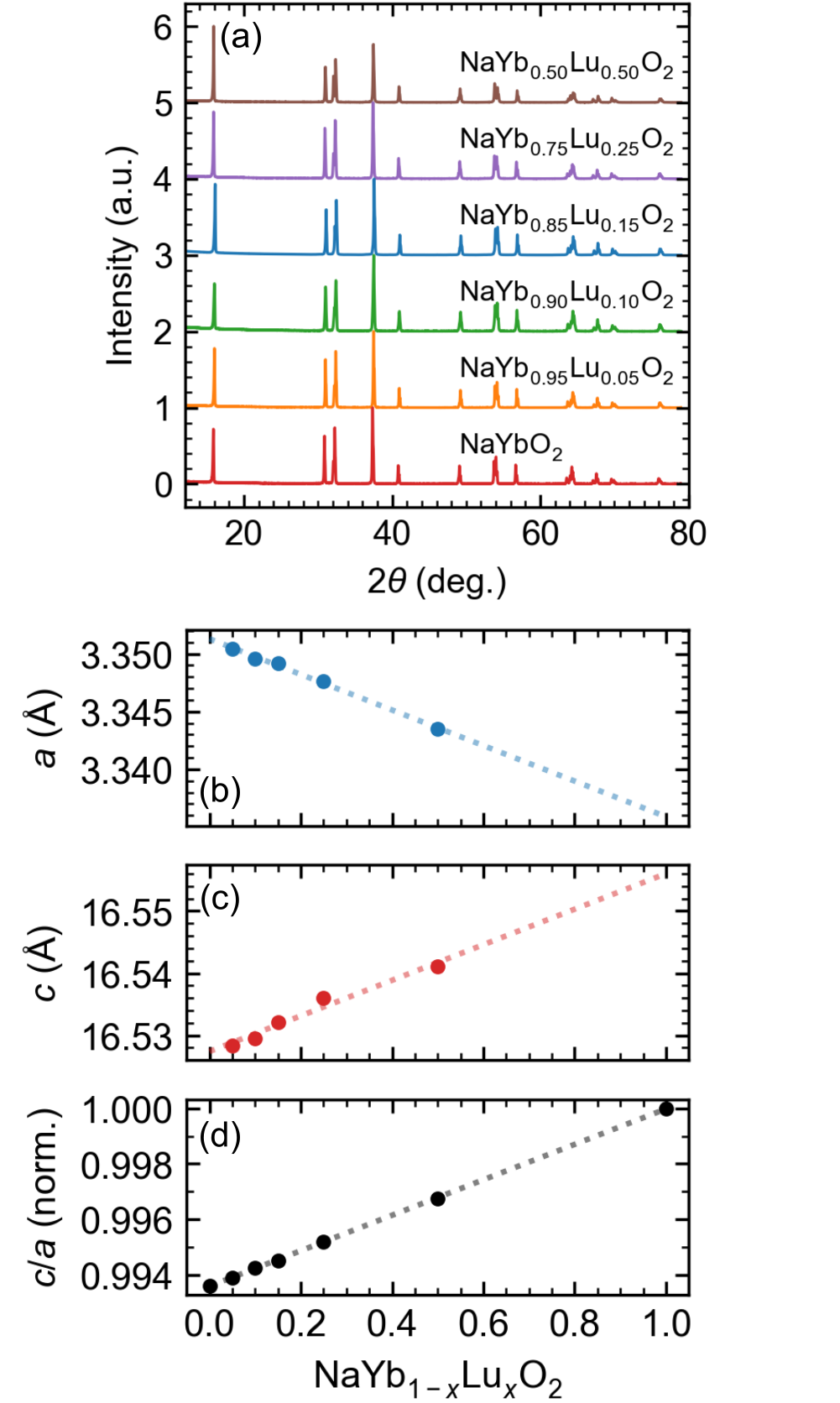}
    \caption{\label{fig:xrd}(a) Laboratory X-ray diffraction patterns collected on samples of NaYb$_{1-x}$Lu$_x$O$_2$. (b) Lattice parameters extracted from Rietveld refinement of profile fits to the diffraction data, with the exception of NaLuO$_2$ which was {adapted} from Ref. \cite{zhang_observation_2024}. The dotted line is provided as a guide to the eye following Vegard's Law. All datasets were collected at room temperature.}
\end{figure}

\begin{figure*}[t]
    \centering
    \includegraphics[width=1.0\linewidth]{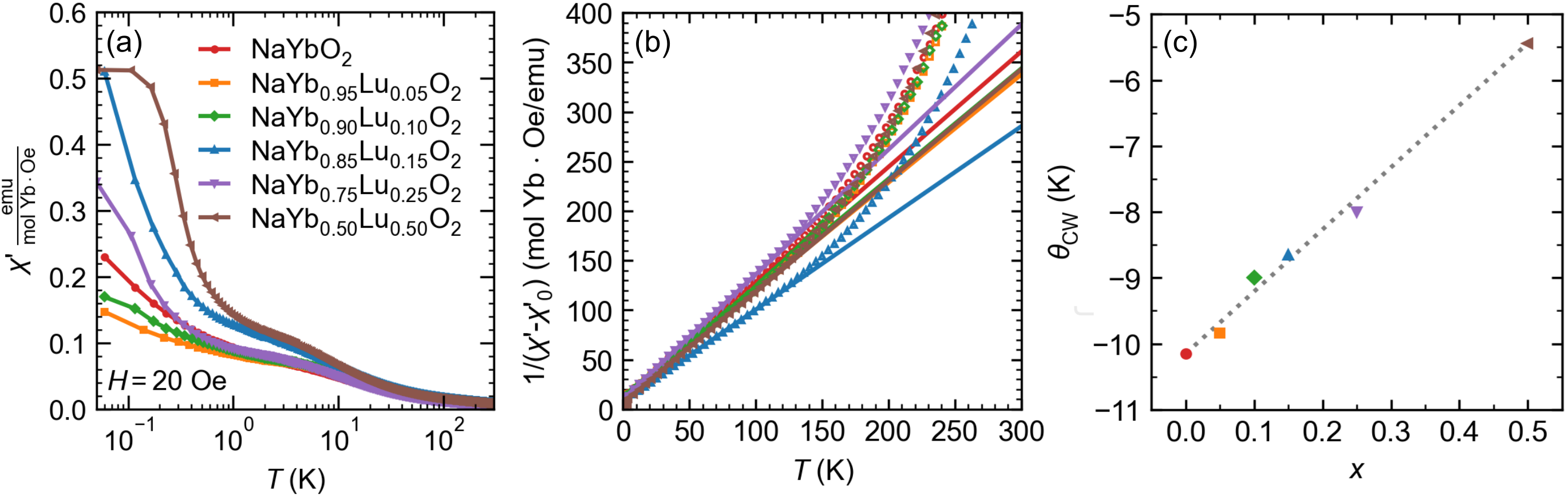}
    \caption{\label{fig:susc} (a) Temperature dependence of the real part of the magnetic susceptibility $\chi'(T)$ for NaYb$_{1-x}$Lu$_x$O$_2$ at $H=20$~Oe. (b) Curie-Weiss fits to the inverse susceptibility and (c) resulting $\theta_\mathrm{CW}$ values as a function of Lu content.}
\end{figure*}

\subsection{Magnetization and heat capacity measurements}

Temperature-dependent dc-magnetization measurements were performed using a Quantum Design Magnetic Property Measurement System (MPMS3) SQUID magnetometer in vibrating-sample (VSM) mode. Polycrystalline NaYb$_{1-x}$Lu$_x$O$_2$ was packed into a polypropylene capsule and mounted into a brass holder.
Temperature-dependent ac-magnetization measurements were performed using a Quantum Design Physical Property Measurement System (PPMS) employing the ac-susceptibility option for the $^3$He/$^4$He dilution refrigerator insert (ACDR). Polycrystalline NaYb$_{1-x}$Lu$_x$O$_2$ was suspended in wax to combat degradation, and a pellet with approximate dimensions 2$\times$2$\times$1~mm$^3$ was adhered to a sapphire sample mounting post with a thin layer of GE varnish. The real part of the measured ac susceptibility was then linearly scaled to the dc magnetization data using a fit to the overlapping temperature region of $T=2$~K to $T=4$~K. 

The heat capacity of NaYb$_{1-x}$Lu$_x$O$_2$ was measured using a 14~T Quantum Design PPMS employing the standard heat capacity option and the corresponding option for the $^3$He/$^4$He dilution refrigerator insert. Polycrystalline NaYb$_{1-x}$Lu$_x$O$_2$  pellets were densified at 
$P=3,275$~bar
in a cold isostatic press and mounted to the sample stage using a small amount of Apezion N grease. To extract the magnetic contribution to the heat capacity, data collected on a nonmagnetic sample of NaLuO$_2$ were scaled and subtracted from each dataset. The data collected in the dilution refrigeration regime were manually scaled by a multiplicative factor to match the high-temperature data in the overlapping temperature region of $T=2$~K to $T=4$~K.

\subsection{$\mu$SR measurements}

The $\mu$SR measurements were performed at the ISIS Pulsed Neutron and Muon Source at Rutherford Appleton Laboratory (UK) using the EMU spectrometer. Pressed disks of polycrystalline NaYb$_{1-x}$Lu$_x$O$_2$, with thickness of approximately 1~mm and diameter of 10~mm were arranged in a close-packed mosaic consisting of 7~disks and mounted in a $^3$He refrigerator. Data were taken between $T=0.25$~K and $T=200$~K and in magnetic fields up to $H=3,000$~G.

\section{Results}

\subsection{Structure and bulk property characterization}

 X-ray diffraction measurements were performed on NaYb$_{1-x}$Lu$_x$O$_2$ with the results shown in Figure 1. The powder patterns for all samples were fit in the R$\overline{3}$m space group and no secondary phases were resolved.  Increasing Lu substitution ($x$) results in a linear decrease of the in-plane lattice parameter $a$ and an increase of the out-of-plane lattice constant $c$, each towards those of NaLuO$_2$. The $c/a$ ratio follows a linear trend across the alloy series, consistent with expectations for Vegard's law.

To characterize the evolution of the magnetic ground state with increasing magnetic disorder, temperature-dependent magnetic susceptibility ($\chi$) measurements were performed at temperatures down to $T=0.05$~K. Figure \ref{fig:susc}(a) shows zero-field cooled $\chi'(T)$ (the real part of the ac-susceptibility), which shows no signatures of long-range order or spin freezing in all samples. 
The frequency dependence of $\chi'(T)$ is presented in Figure S2, demonstrating little to no variation for frequencies between 10~Hz and 10~kHz.
Below $T=1$~K, a rise in $\chi'$ appears, consistent with a small fraction of free Yb$^{3+}$ moments which is expected to be sample-dependent \cite{bordelon_field-tunable_2019}.  
{The amount of impurity spins was estimated via a fit of the field-dependent magnetization to a two-component model and demonstrates no systematic dependence on the strength of the dilution (Fig. S4).}
The temperature-dependent susceptibility data between $T=20$~K and $T=100$~K were fit to a Curie-Weiss model of the form
\begin{equation}
    \frac{1}{\chi-\chi_0} = \left(\frac{C}{T-\theta_\mathrm{CW}}\right)^{-1}
\end{equation}
\noindent
where $C$ is the Curie constant, $T$ is the temperature, $\chi_0$ is the temperature-independent contribution to the susceptibility, and $\theta_\mathrm{CW}$ is the Curie-Weiss temperature [Fig \ref{fig:susc}(b)].
From these fits, we find that the effective moment per Yb ion $\mu_\mathrm{eff}=\sqrt{8C}$ remains relatively unchanged across the series, whereas $\theta_\mathrm{CW}$ evolves linearly from $\theta_\mathrm{CW}\approx-10$~K for pure NaYbO$_2$ to $\theta_\mathrm{CW}\approx-5$~K for $x=0.50$, consistent with a picture of nonmagnetic site disorder which disrupts the exchange network in a classical Heisenberg model \cite{RevModPhys.58.801} [Fig. \ref{fig:susc}(c)]. 
These results agree with $\mu_\mathrm{eff}$ and $\theta_\mathrm{CW}$ values reported elsewhere \cite{zhang_observation_2024}. 
This higher temperature assessment of the mean-field exchange agrees well with field-dependent magnetization measurements (Fig. S3), and the linear trend of both with $x$ supports the spatial homogeneity of the diluted triangular rare-earth network across the sample series, despite varying levels of free moment disorder evidenced at low temperatures.

\begin{figure*}[t]
    \centering
    \includegraphics[width=1.0\linewidth]{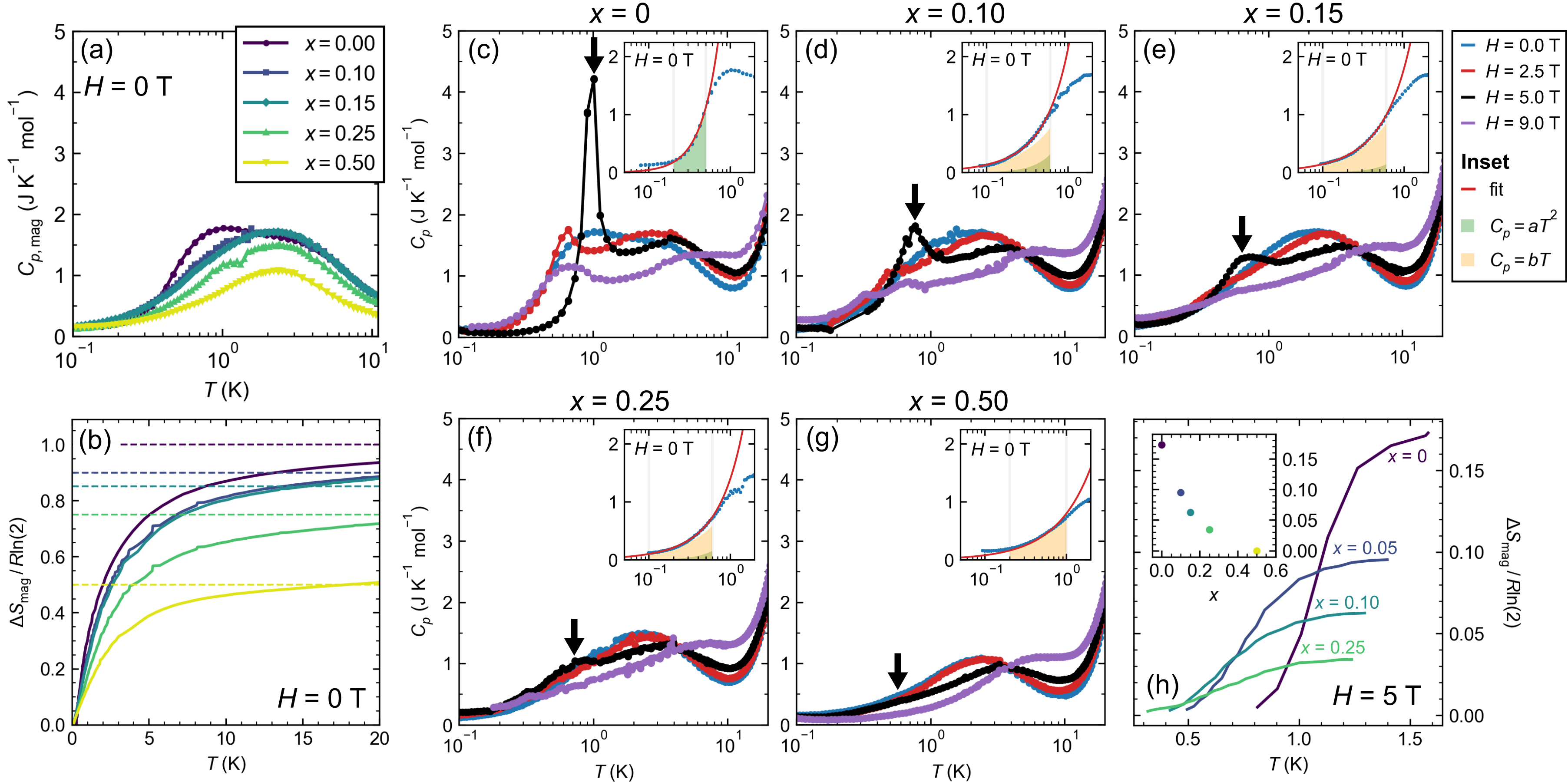}
    \caption{\label{fig:hc} (a) Magnetic contribution to the zero-field heat capacity at low temperature for each composition. (b) Integrated zero-field entropy release in units of $R\ln(2)$. Dashed lines mark the expected entropy release for each composition. (c-g) Field-dependence of the total heat capacity for select compositions. A black arrow marks the $\lambda$-anomaly associated with the field-induced two-$\mathbf{q}$ magnetic order in the $H=5~\mathrm{T}$ data. Insets: Low-temperature fits to the zero-field magnetic contribution to the heat capacity using the model $C_p(T) = aT^2+bT$. Shaded regions represent the individual contributions from the quadratic and linear terms in the model fit. (h) Integrated magnetic entropy release for the $\lambda$-anomaly at $H=5$~T. Inset: Magnetic entropy release at $H=5$~T as a function of composition.
    }
\end{figure*}

\begin{figure*}[t]
    \centering
    \includegraphics[width=1.0\linewidth]{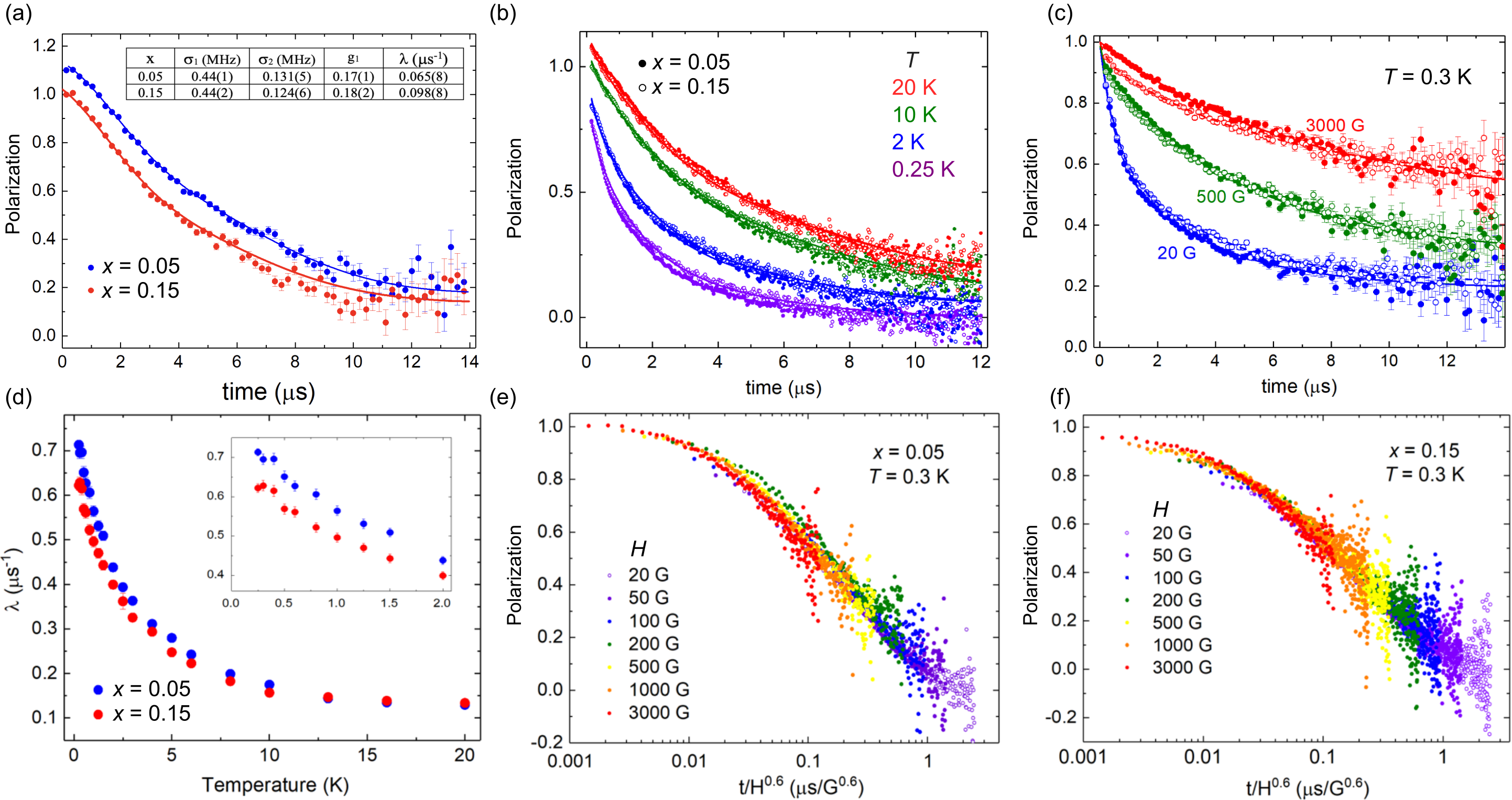}
    \caption{\label{fig:muons123456}     (a) Zero field depolarization at temperatures of 195~K ($x = 0.05$) and 130~K ($x = 0.15$). 
    Curves are offset for clarity. 
    (b) Select depolarization data in zero magnetic field for $x = 0.05$ (filled circles) and 0.15 (open circles). 
    Curves are offset for clarity. 
    (c) Select depolarization data in applied longitudinal magnetic fields at $T = 0.3~K$ for $x = 0.05$ (filled circles) and 0.15 (open circles). 
    The solid lines in each panel (a-c) are fits to Eq. \ref{eq:polarization}.
    (d) Values for the depolarization rate versus temperature in zero magnetic field as extracted from fits to Eq. \ref{eq:polarization}. Inset: expanded view of the low-temperature variation of $\lambda$. 
    (e) Longitudinal magnetic field scaling data for $x = 0.05$ and (f) $x = 0.15$.}
\end{figure*}

To further characterize the evolution of magnetic correlations, heat capacity measurements were performed as a function of Lu concentration $x$.  Under zero field, all samples show no signatures of order down to 50~mK for samples with Lu compositions $x\leq0.50$ [Fig. \ref{fig:hc}(a)] with only broad peaks signifying the formation of short-range correlations. Whereas in pure NaYbO$_2$ a broad, two-peak magnetic feature emerges at low temperature, for $x\geq0.10$ we find that the lower temperature peak is suppressed and the corresponding entropy is transferred to the broader, higher temperature peak.   Integrating over both peaks shows that the expected magnetic entropy is preserved and scales with the diluted fraction of Yb moments, as shown in Fig. \ref{fig:hc}(b).

Given that the g-tensor values and anisotropy are known to be only weakly perturbed upon Lu substitution in NaYb$_{1-x}$Lu$_x$O$_2$ \cite{zhang_observation_2024}, the relative stability of the field-induced up-up-down state can be probed via heat capacity measurements under applied field.  At 5 T, the powder-averaged field is known to drive a global transition into this state in pristine NaYbO$_2$ \cite{bordelon_field-tunable_2019}, which is signified in $C_p(T)$ via a sharp peak that appears as an apparent condensation of the broad, lower-temperature zero-field peak. Fig. \ref{fig:hc}(c) shows this peak and its evolution upon magnetic dilution, where the field-induced order is smoothly suppressed with increasing Lu content [Fig. \ref{fig:hc}(c-g)]. In the $x=0.50$ sample, this ordering transition is absent, consistent with the expected site percolation limit for a 2D triangular lattice network. Figure \ref{fig:hc}(h) shows the estimated entropy associated with this field-induced order as a fraction of the total, $R\ln(2)$.  The inset shows the corresponding suppression of this fraction under Lu substitution.

Analyzing the low-temperature $C_p(T)$ data at zero-field, the evolution of magnetic excitations below the lowest temperature broad peak was previously fit to a $T^2$ power law \cite{bordelon_field-tunable_2019}. With increasing Lu content, this remnant low-temperature heat capacity trends toward a linear temperature dependence as shown in the insets of Figs. \ref{fig:hc}(c-g). To parametrize this, data were fit to a simple two-component model $C_p(T) = aT^2+bT$ with the trade-off between each component shown as the shaded regions in the inset. A rapid shift to linear $C_p(T)$ is seen by $x=0.15$, reflecting a qualitative change in the low-temperature magnetic excitations.

\subsection{Muon spin relaxation}

In order to probe the change-over in the low-energy magnetic fluctuations more deeply, $\mu$SR measurements were employed as a definitive probe of the local field within two NaYb$_{1-x}$Lu$_x$O$_2$ samples of compositions $x=0.05$ and $x=0.15$.
In Fig. \ref{fig:muons123456}(a), muon depolarization data are shown at high temperatures for both samples. Nearly identical results appear for both $x=0.05$ and $x=0.15$ compositions, characterized by a two-component response (excluding the background signal). At high temperatures, the Yb$^{3+}$ paramagnetic fluctuations are very fast and mostly outside the bandwidth of the muon response window, and the depolarization is primarily due to $^{23}$Na $I=3/2$ nuclei. The high temperature data are well described by
\begin{equation}
    P(t)=(1-f_\mathrm{bg})\exp\left[-(\lambda_\mathrm{HT}t)^\beta\right]\sum_{i=1}^2 g_iG_\mathrm{KT}(\sigma_i;t)+f_\mathrm{bg}.
    \label{eq:polarization}
\end{equation}

\noindent 
$G_\mathrm{KT}(\sigma_i;t) = \frac{2}{3}(1-\sigma^2t^2)\exp(-\frac{\sigma^2t^2}{2})+\frac{1}{3}$ is the Gaussian Kubo-Toyabe depolarization function \cite{amato_introduction_2024}, typical of a random quasistatic Gaussian distribution of magnetic moments, and $\sigma=\gamma_\mu\Delta$, with $\Delta$ being the 2$^\mathrm{nd}$ moment of the Gaussian field distribution and $\gamma_\mu = 0.0851$~MHz/G is the muon gyromagnetic ratio. $f_\mathrm{bg}$ is the fraction of muons landing outside the magnetic samples, being 0.10 for both samples, and $g_i$ are the relative amplitudes of the two magnetic contributions with $g_1+g_2=1$. 

The exponential decay is depolarization due to the tail end of the Yb fluctuation spectrum \cite{exponential_decay_note}. The exponent $\beta=0.99(2)$, \textit{i.e.}, pure exponential behavior, is observed as expected for uncorrelated paramagnetic fluctuations. Values for the other fit parameters for the two concentrations are shown in the inset of Fig. \ref{fig:muons123456}(a). About 17\% of the muons experience stronger nuclear depolarization, suggesting that they are stopping at positions closer to the Na layer.

At lower temperatures, the Yb$^{3+}$ moments dominate the zero-field depolarization, and exponential-like depolarization is observed down to 0.25~K [Fig. \ref{fig:muons123456}(b)], demonstrating a lack of magnetic ordering or spin freezing. If the magnetism were quasi-static, application of a strong magnetic field along the initial spin polarization direction would dominate the local magnetic field and suppress depolarization. More quantitatively, in the quasi-static case suppression of relaxation is nearly complete if the longitudinal field $B_\mathrm{LF}\approx10 \lambda(0)/\gamma_\mu \equiv B_c$ where $\lambda(0)$ is the low temperature relaxation rate in zero field \cite{exponential_decay_note}; visual inspection of the data in Fig. \ref{fig:muons123456}(b) shows $\lambda(0) \approx 2~\mathrm{\mu s}^{-1}$ yielding $B_c\approx240~\mathrm{G}$. In Fig. \ref{fig:muons123456}(c) some typical depolarization curves taken at $T=0.3~\mathrm{K}$ and various magnetic field values are shown. Significant depolarization remains evident up to 3000~G, demonstrating relaxation via primarily dynamic processes. 

Now turning to a more quantitative analysis, the zero-field depolarization for NaYbO$_2$ \cite{PhysRevB.100.144432} was previously modeled as the sum of two exponential decays, each proposed to result from muons stopping at two different crystalline sites. However, while this model provides a good phenomenological fit to the zero-field data in both Ref. \cite{PhysRevB.100.144432} ($x=0$) and for the present data with very similar parameter values (see the Supplemental Material \cite{supp_info_note}), it presents some inconsistencies as well: (1) Below 10~K the two components are in a ratio of 1:1, whereas the high-temperature data show two stopping sites in ratio of approximately 5:1; (2) The temperature and field dependencies of the two extracted relaxation rates are qualitatively different (Fig. S5), which is not expected for muons at two different lattice sites being depolarized by the same magnetic fluctuations. An alternative explanation assuming two magnetically inequivalent phases, as in Ref. \cite{zhu_fluctuating_2023}, is also problematic: it is unlikely that the second phase nucleates out of the first with volume fraction of 0.5 for three different samples with different levels of Lu doping. 

Therefore, we consider a single stretched exponential function, as in Eq. \ref{eq:polarization}, to describe the depolarization of muons stopped at the two muon sites determined at high temperature. Note that in the Supplemental Information provided in Ref. \cite{PhysRevB.100.144432} the 
authors 
state that their data for NaYbO$_2$ could also have been described by a stretched exponential function. For our low-temperature fits, the values of $\sigma_1$, $\sigma_2$, $g_1$, and $f_\mathrm{bg}$ were fixed to their high-temperature values. Additionally, for temperatures at and below 20~K the exponent was determined to be $\beta\approx0.65$, consistent with a model demonstrating a complex fluctuation spectrum beyond the single correlation time approximation, and so was fixed at this value. The resulting fit curves are shown as solid lines in Fig. \ref{fig:muons123456}(b), with the temperature variation of $\lambda$ shown in Fig. \ref{fig:muons123456}(d). The rise in depolarization rate below $\approx$10~K is interpreted as the onset of correlated fluctuations in this system, and it appears to decrease only slightly upon increasing $x$ from 0.05 to 0.15.

\section{Discussion}

{Our heat capacity measurements reveal that} the zero-field state is only weakly perturbed with Lu substitution, with the main impact being a redistribution of magnetic entropy upward in temperature and the elimination of the lowest temperature, broad peak in $C_p(T)$.
In the pristine system, this low-T peak fully condenses into an up-up-down state under applied field, and data show that this field-induced state is suppressed at the $x=0.5$ percolation limit. 
The rapid suppression of the field-induced collinear order and its greater sensitivity to dilution is consistent with expectations for nonmagnetic disorder on the triangular lattice \cite{maryasin_triangular_2013}.

Prior to that, however, the low-temperature power law behavior of $C_p(T)$ changes under light impurity substitution and suggests a change in the low-frequency magnetic fluctuations driven by the introduction of disorder.  Whereas in the pure NaYbO$_2$ compound, gapless spin excitations from an underlying Dirac spin liquid state on the triangular lattice \cite{savary_quantum_2016} may generate the $T^{2}$ behavior, the evolution toward a linear-$T$ behavior is potentially indicative of very low frequency glassy dynamics induced via nonmagnetic disorder---although neither low-temperature susceptibility nor $\mu$SR measurements detect conventional freezing.

$\mu$SR measurements instead unveil the presence of correlated spin fluctuations in impurity substituted samples with a spectrum reminiscent of the pristine NaYbO$_2$ compound. 
The presence of correlated spin fluctuations implies a scaling behavior in applied longitudinal fields $H$ such that the time-dependent polarization has the form \cite{PhysRevB.64.054403, keren_muons_2004}

\begin{equation}
    P(H,t)=P(t/H^\gamma)
    \label{eq:scaling}
\end{equation}

\noindent
where the exponent $\gamma$ depends on the detailed form of the spin correlation function $S(t)$. The data for $x=0.05$ and $x=0.15$ (Fig. \ref{fig:muons123456}(e,f)) are well-described over more than three orders of magnitude by Eq. \ref{eq:scaling} assuming an exponent of $\gamma=0.6$. This suggests a power-law prefactor to the spin correlation function such that $S(t)\propto (\frac{\tau}{t})^x\exp(-\nu t)$, with $\tau$ being a short-time cut-off, and $\gamma=1-x$ \cite{keren_muons_2004}. Similar exponents $\gamma<1$ are found in other spin-liquid candidates \cite{PhysRevB.105.L180402, PhysRevB.110.134412, yang_muon_2024}, although these are significantly different from $\gamma=1.75$ as found for the triangular lattice magnet NaRuO$_2$ \cite{ortiz_quantum_2023}, suggesting very different dynamics are in play for that system. {A fit of the longitudinal field dependence of the relaxation rates using both an early and late time cut-off reveals timescales much slower than paramagnetic estimates (Fig. S6), and a value of $x=0.375$ in agreement with the $\gamma$ obtained from the scaling analysis (see the Supplemental Material \cite{supp_info_note}). 
The temperature dependence of the relaxation rate is markedly different from that in pristine NaYbO$_2$ \cite{PhysRevB.100.144432}, which featuers a low-temperature plateau below $T\approx3$~K. This behavior is likely shifted to lower temperatures upon magnetic dilution for the $x=0.05$ and $x=0.15$ samples measured here, and is closer to the temperature scale of the plateau observed in the muon spin relaxation rate in YbMgGaO$_4$ \cite{li_MuonSpinRelaxation_2016}. However, non-saturating behavior has been observed in CeMgAl$_{11}$O$_{19}$ \cite{cao_U1DiracQuantum_2025a}, and thus lower temperature measurements are necessary to better qualify the behavior of magnetically diluted NaYbO$_2$ in this context.}

Our results complement similar studies exploring the impact of magnetic dilution in quantum disordered NaYb$X_2$ systems. In NaYb$_{1-x}$Lu$_x$Se$_2$ \cite{pritchard_cairns_tracking_2022} susceptibility and heat capacity data show the formation of isolated dimers in the dilute limit, which progressively evolve into a correlated state as NaYbSe$_2$ is approached.  In the dilute limit, the local dimers can be precisely characterized and the singlet-triplet transition driven via an external field \cite{zhang_method_2025}.  A similar evolution of dimers and a robust parent quantum disordered state was also reported in NaYb$_{1-x}$Lu$_x$S$_2$ \cite{PhysRevMaterials.6.046201}.  In this and other studies, the local Yb-Yb exchange and Yb$^{3+}$ anisotropies are only weakly perturbed upon site dilution. 

Finally, a recent study of NaYb$_{1-x}$Lu$_x$O$_2$ reports similar behaviors in terms of dimer formation in the dilute limit \cite{zhang_observation_2024} and the resilience of anisotropies and local exchange under Lu substitution.  Our data further this picture by probing the evolution of magnetic dynamics and correlations close to the pristine compound and toward the percolation threshold where an extensive entangled state survives. 

\section{Conclusions}

{Our combined results characterizing the evolution of the magnetic state in NaYb$_{1-x}$Lu$_x$O$_2$ reveal a quantum disordered state that is robust to magnetic dilution up to the percolation threshold.
Bulk thermodynamic measurements show no obvious signs of classical order or spin freezing up to 50\% site dilution.
The robust low-frequency, low-temperature dynamics resolved by $\mu$SR measurements, even in the presence of 15$\%$ site dilution, and the preclusion of a magnetically heterogeneous response places stringent constraints on microscopic models for quantum disorder in NaYbO$_2$. Future $\mu$SR measurements at higher dilution levels and at lower temperatures will further illuminate the nature of the ground state in this system.}

\section{Acknowledgments}

The authors acknowledge various forms of support from Mitchell Bordelon, Paul Sarte, Caeli Benyacko, Dibyata Rout, Roland Yin, and Casandra Gomez Alvarado.
This work was supported by the US Department of Energy (DOE), Office of Basic Energy Sciences, Division of Materials Sciences and Engineering under Grant No. DE-SC0017752.
S.J.G.A. acknowledges the additional financial support from the National Science Foundation Graduate Research Fellowship under Grant No. 1650114.
Work by B.R.O. was supported by the U.S. Department of Energy (DOE), Office of Science, Basic Energy Sciences (BES), Materials Sciences and Engineering Division.
This research made use of the shared facilities of the NSF Materials Research Science and Engineering Center at UC Santa Barbara (DMR-2308708). Experiments at the ISIS Neutron and Muon Source were supported by beam time allocation RB2310423 from the Science and Technology Facilities Council.

\section{Author declarations}
\subsection{Conflict of interest}
The authors have no conflicts to disclose.	

\section{Data availability}
The data that support the findings of this study are openly available \cite{zenodo}. Data collected at ISIS can be found at Ref. \cite{isis_doi}.

\bibliography{bibliography}

\newpage
\onecolumngrid
\newpage
\part*{Supplemental Information}

\renewcommand{\arraystretch}{0.6} 
\renewcommand{\thetable}{S\arabic{table}}
\renewcommand{\thefigure}{S\arabic{figure}}
\setcounter{figure}{0}    

Our initial analysis followed that reported in Ref. [1] for pure NaYbO$_2$

\begin{equation}
    P(t) = (1-f_\mathrm{bg})\left[f_1\exp(-\lambda_1t)+f_2\exp(-\lambda_2t)\right]+f_\mathrm{bg}
    \label{eq:polarization_supp}
\end{equation}

\noindent
with $f_1+f_2=1$ and $f_\mathrm{bg}$ fixed at the values determined from the high-temperature data. Some representative depolarization data with fits are shown in Fig. 4(a).
The temperature dependence of $\lambda_1$ and $\lambda_2$ are shown in Fig. S\ref{fig:muonsS12}(a). In order to accommodate the crossover in the magnetic profile from nuclear-dominated to electronic-dominated, $f_1$ was treated as a temperature-dependent fit parameter, and is shown in the inset of Fig. \ref{fig:muonsS12}(a); at 20~K the fast response accounts for 30\% of our magnetic signal, rising to near 50\% below. It is clear that the magnetic response is nearly identical for $x=0.05$ and $0.15$ samples. Our fit parameters are essentially the same as those reported in Ref. [1], apart from a roughly factor of 2 difference in the slow rate $\lambda_2$, which could be accounted for by differences in background contributions and sample quality. We note that the fraction $f_1$ reported in Ref. [1] is 0.5, but data is only presented up to 10~K.

We fit the LF data at $T=0.3$~K for the $x=0.05$ sample to Eq. \ref{eq:polarization_supp}, assuming that $f_1$ and $f_2$ remain at their zero-field values. In Fig. \ref{fig:muonsS12}(b) we show the longitudinal field dependence of $\lambda_1$ and $\lambda_2$, both normalized to their zero-field values. For depolarization via fluctuations the field dependence of the relaxation rate normalized to its zero-field value depends only on the details of the fluctuation spectrum [2]. Hence, for a magnetically homogeneous material (\textit{i.e.}, no phase separation), muons stopping in different lattice sites should have normalized rates with identical longitudinal field dependencies, \textit{i.e.}, they should all lie on the same curve. Inspection of Fig. \ref{fig:muonsS12}(b) shows that is not the case here, and we conclude that NaYb$_{1-x}$Lu$_x$O$_2$ ($x\leq0.15$) cannot be described by a simple model of two stopping sites in a magnetically homogeneous material.

We also fit the LF dependence of the muon spin relaxation rate $\lambda$ to 
\begin{equation}
    \label{eq:rate_vs_field}
    \lambda(H) = 2 \Delta^2 \tau^x \int_0^\infty t^{-x} e^{-\nu t} \cos(2 \pi \mu_0 \gamma_\mu H t)~dt
\end{equation}

\noindent following the treatment in Ref. [3], where $\Delta$ is the width of a Gaussian distribution of local magnetic fields, $\tau$ and $1/\nu$ are early and late time cutoffs, respectively, and $\gamma_\mu$ is the $\mu^+$ gyromagnetic ratio. The results are presented in Fig. S6(a), and reveal a low frequency consistent with that found in YbMgGaO$_4$ (signficantly lower than paramagnetic estimates). We also find a value of $x=0.375$ consistent with our value of $\gamma = 0.6$. The value of $\Delta$ is lower than expected from paramagnetic estimates, though this may be a consequence of the quantum correlations in the ground state.

\begin{figure*}[t]
    \centering
    \includegraphics[width=0.8\linewidth]{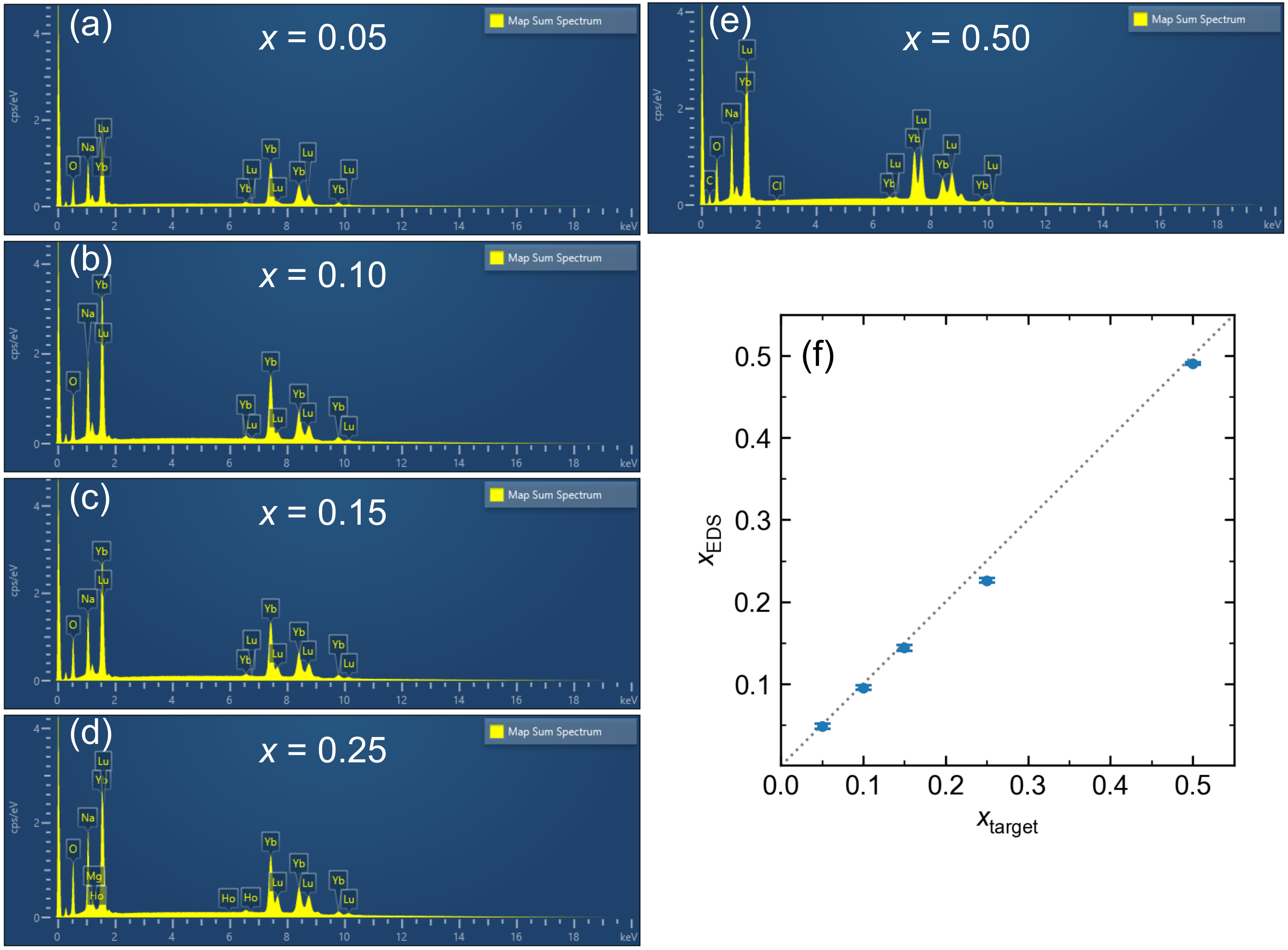}
    \caption{\label{fig:eds} (a--e) Energy-dispersive spectra for Lu-substituted samples. (f) Comparison of measured vs. nominal Lu stoichiometry $x$. Error bars represent one standard deviation.}
\end{figure*}

\begin{figure*}[t]
    \centering
    \includegraphics[width=1.0\linewidth]{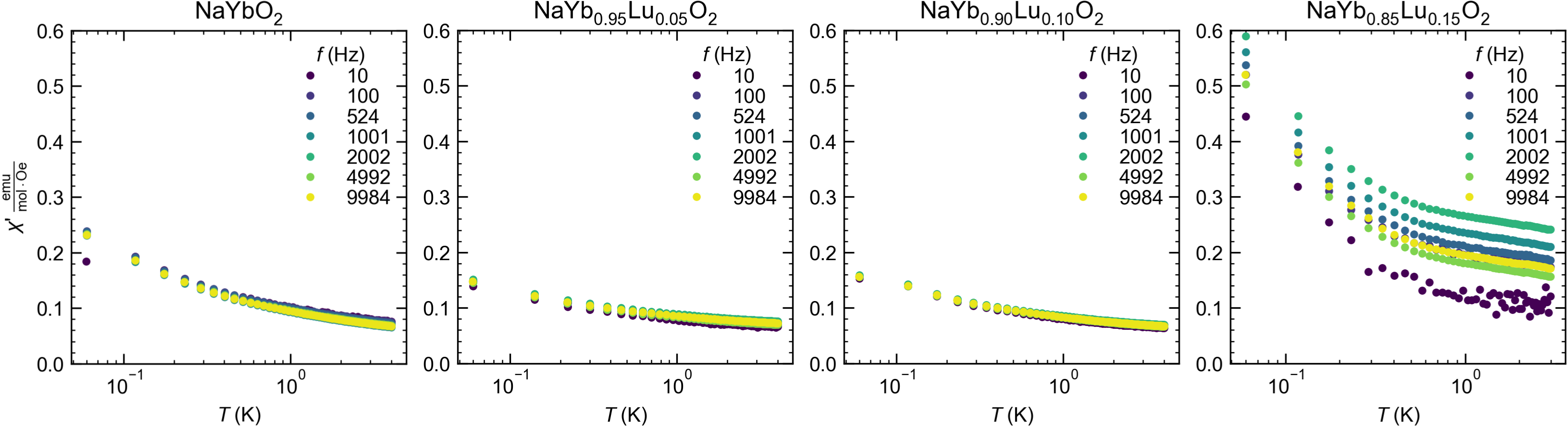}
    \caption{\label{fig:freq_dependence} Real part of the temperature-dependent ac magnetic susceptibility on select samples from the dilution series as a function of frequency.}
\end{figure*}

\begin{figure*}[t]
    \centering
    \includegraphics[width=0.35\linewidth]{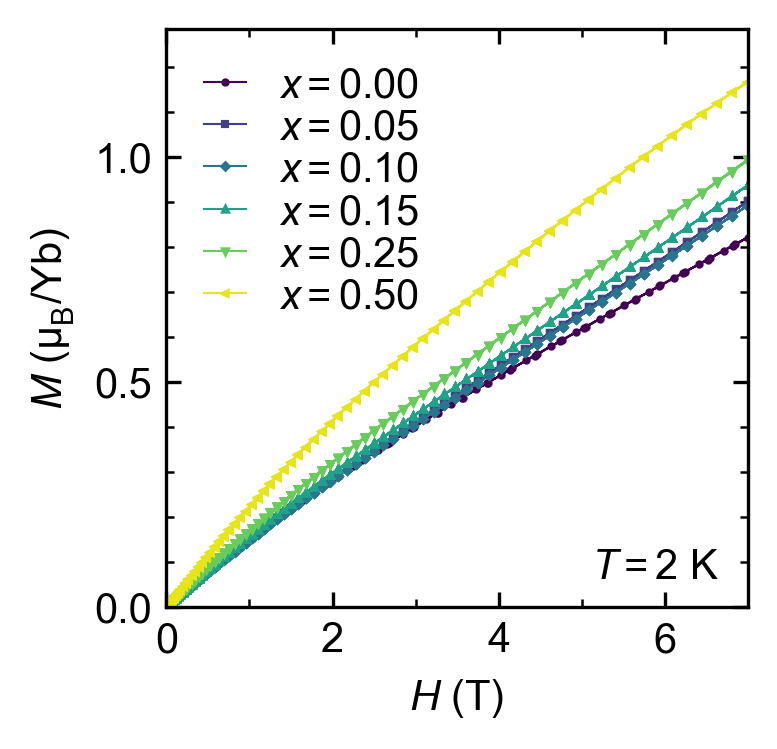}
    \caption{\label{fig:mh_curves}Field-dependent magnetization for samples across the dilution series collected at $T=2$~K.
    }
\end{figure*}

\begin{figure*}[t]
    \centering
    \includegraphics[width=0.8\linewidth]{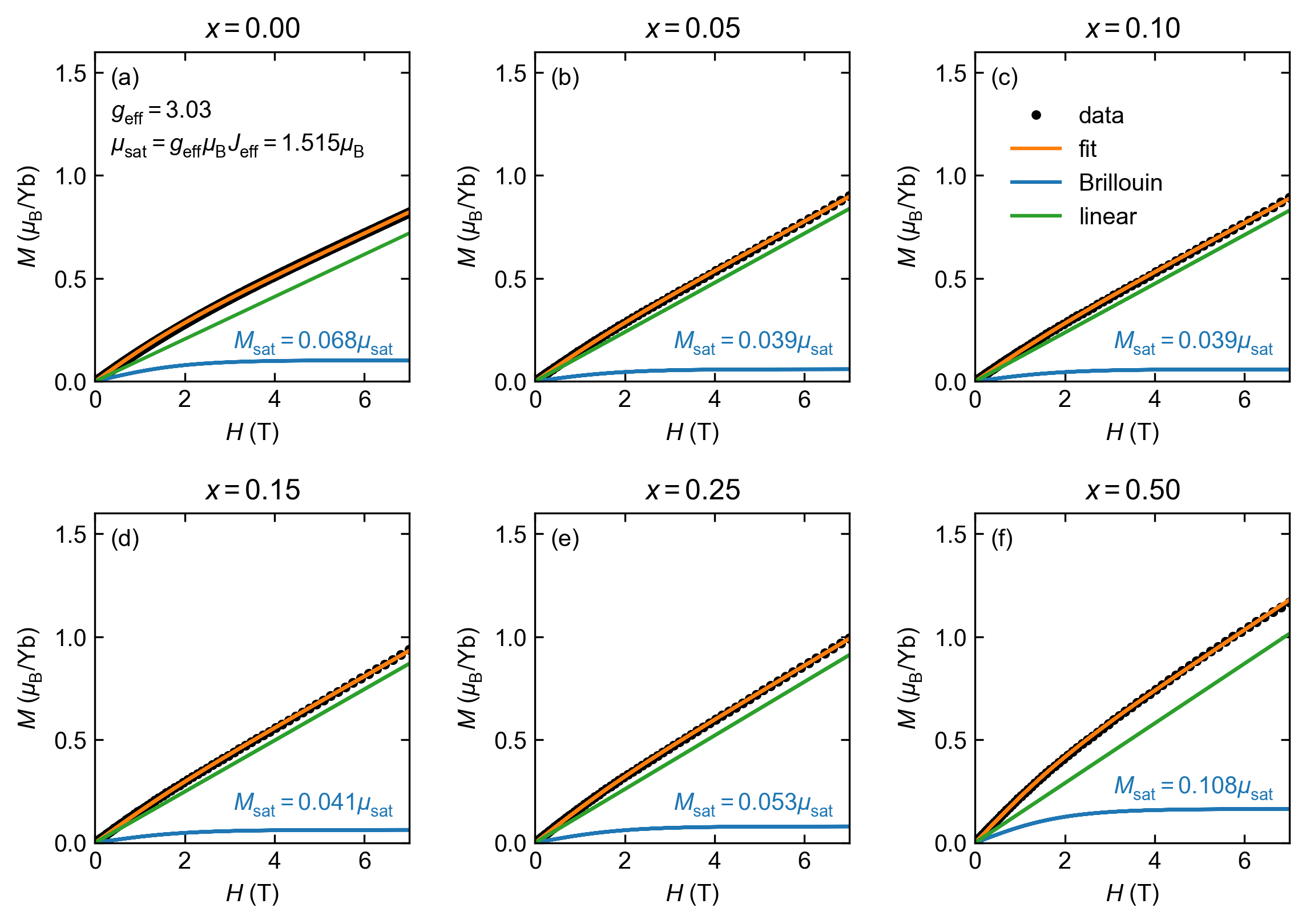}
    \caption{\label{fig:mu_sat} Field-dependent magnetization at $T=2$~K and fits to a two-component model for (a) $x=0$, (b) $x=0.05$, (c) $x=0.10$, (d) $x=0.15$, (e) $x=0.25$, and (f) $x=0.50$. The data was fit to a Brillouin function assuming the average $g$-factor in as reported in Ref. [4], and $J_\mathrm{eff}=1/2$ moments with a variable volume fraction, combined with a linear response due to the strong antiferromagnetic exchange field of the host network of moments. The fraction of free spins determined for each composition is denoted in each panel.
    }
\end{figure*}

\begin{figure*}[t]
    \centering
    \includegraphics[width=1.0\linewidth]{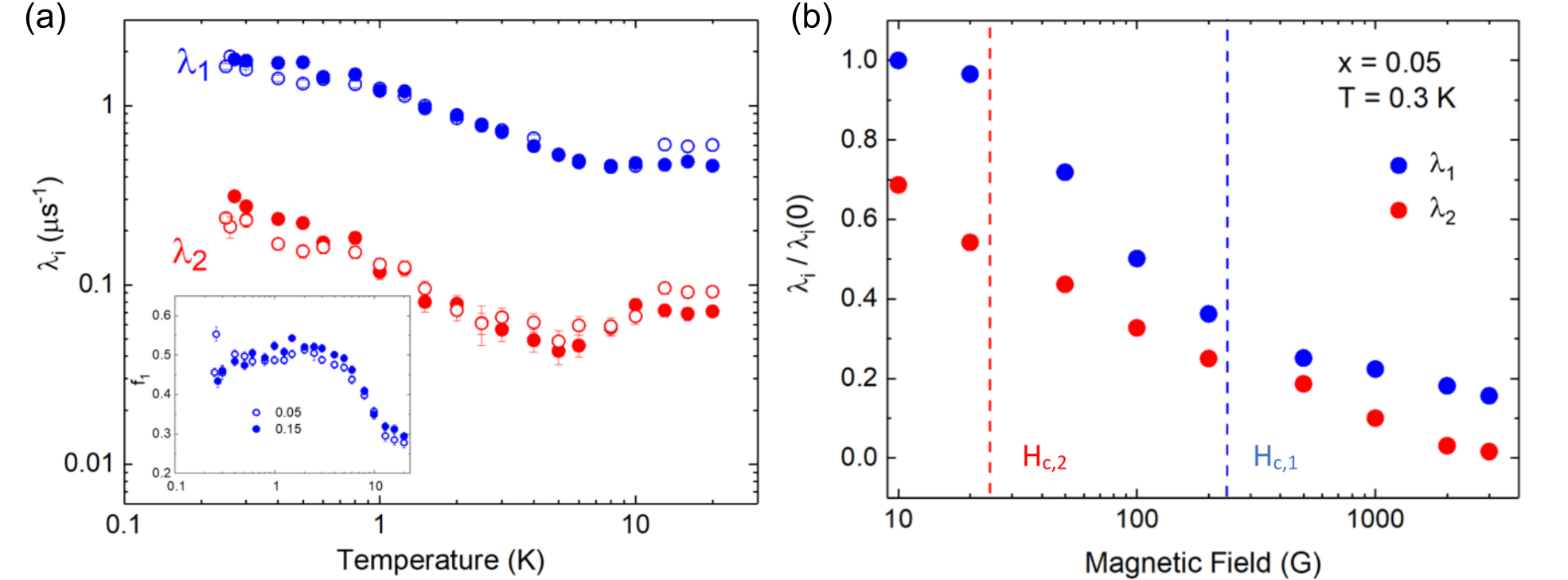}
    \caption{\label{fig:muonsS12} (a) Temperature dependence of the relaxation rates from Eq. 2 in zero magnetic field for x = 0.05 (filled circles) and 0.15 (open circles). Inset: depolarization occurs via static fields (main text). (b) Normalized relaxation rates vs. magnetic field. $H_{c,1}$ and $H_{c,2}$ are the field values at which complete suppression of relxation is expected if depolarization occurs via static fields (main text).}
\end{figure*}

\begin{figure*}[t]
    \centering
    \includegraphics[width=1.0\linewidth]{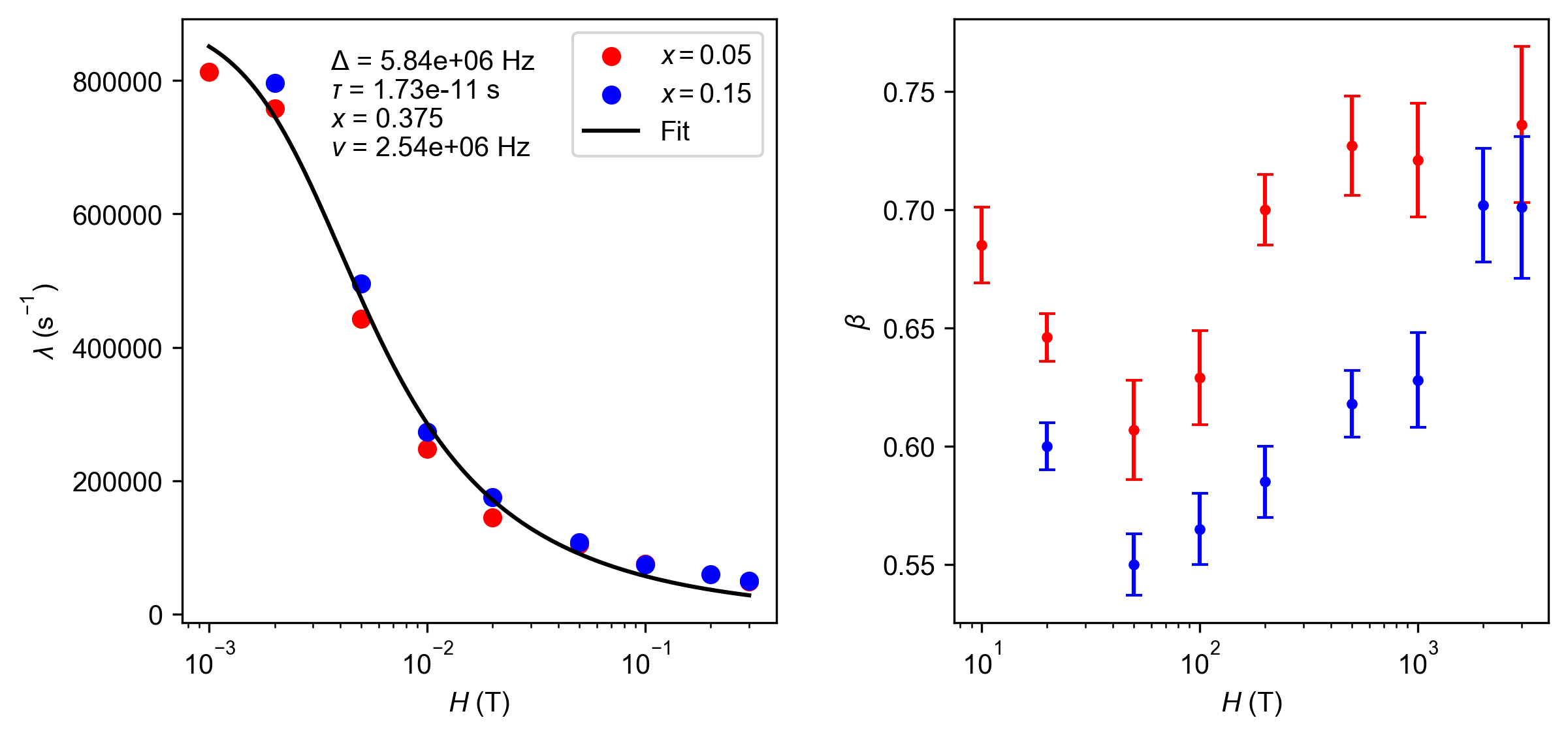}
    \caption{\label{fig:rate_vs_field} (a) Longitudinal field dependence of the muon spin relaxation rate $\lambda$ and (b) the stretching exponents $\beta$ at $T=0.3$~K. The solid line in (a) represents the fit to Eq. \ref{eq:rate_vs_field}, with refined parameters reported in the figure panel.}
\end{figure*}


[1] L. Ding, P. Manuel, S. Bachus, F. Grußler, P. Gegenwart, J. Singleton, R. D. Johnson, H. C. Walker, D. T. Adroja, A. D. Hillier, and A. A.
Tsirlin, Gapless spin-liquid state in the structurally disorder-free triangular antiferromagnet NaYbO$_2$, \textit{Phys. Rev. B} \textbf{100}, 144432 (2019).
[2] A. Keren, Muons as probes of dynamical spin fluctuations: some new aspects, \textit{J. Phys. Condens. Matter} \textbf{16}, S4603 (2004).
[3] Y. Li, D. Adroja, P. K. Biswas, P. J. Baker, Q. Zhang, J. Liu, A. A. Tsirlin, P. Gegenwart, and Q. Zhang, Muon Spin Relaxation Evidence
for the $U(1)$ Quantum Spin-Liquid Ground State in the Triangular Antiferromagnet YbMgGaO$_4$, \textit{Phys. Rev. Lett.} \textbf{117}, 097201 (2016).
[4] M. M. Bordelon, E. Kenney, C. Liu, T. Hogan, L. Posthuma, M. Kavand, Y. Lyu, M. Sherwin, N. P. Butch, C. Brown, M. J. Graf, L. Balents,
and S. D. Wilson, Field-tunable quantum disordered ground state in the triangular-lattice antiferromagnet NaYbO$_2$, \textit{Nat. Phys.} \textbf{15}, 1058
(2019).

\end{document}